# Laser Induced Magnetization Reversal for Detection in Optical Interconnects

Zubair Al Azim, Xuanyao Fong, Thomas Ostler, Roy Chantrell, and Kaushik Roy

*Abstract*— **Optical interconnect has emerged as the front-runner to replace electrical interconnect especially for off-chip communication. However, a major drawback with optical interconnects is the need for photodetectors and amplifiers at the receiver, implemented usually by direct bandgap semiconductors and analog CMOS circuits, leading to large energy consumption and slow operating time. In this article, we propose a new optical interconnect architecture that uses a magnetic tunnel junction (MTJ) at the receiver side that is switched by femtosecond laser pulses. The state of the MTJ can be sensed using simple digital CMOS latches, resulting in significant improvement in energy consumption. Moreover, magnetization in the MTJ can be switched on the picoseconds time-scale and our design can operate at a speed of 5 Gbits/sec for a single link.**

## I. INTRODUCTION

OPTICAL interconnects have been proposed as the technology for replacing Cu based global interconnects with Cu wires due to their superior signal transmission characteristics in the nano-regime [1], [2]. However, a major obstacle to the optical scheme is the need for optical-to-electrical signal conversion. Direct bandgap semiconductor based photodetectors are usually needed to perform this conversion and power hungry amplifiers are needed to amplify the electrical signals to levels suitable for digital logic [1], [3]. The slow response time, necessity of engineering and integrating suitable direct-bandgap materials with the CMOS process, and the challenge of maintaining high quantum efficiency and low thermal noise in the photodetectors have severely limited on-chip optical communication [2]. In this work, we present a new scheme for using optical signals directly with standard CMOS circuits by directly switching a Magnetic Tunnel Junction (MTJ) using ultrashort laser pulses. Data is transmitted using femtosecond laser pulses which induce magnetization reversal in the MTJ in the receiver. The MTJ state is then sensed using simple digital CMOS components. Our proposed scheme does not require photo-diodes and subsequent amplifiers, thus offering almost 40% improvement from the current technology value for the target speed of 5 Gbits/sec for a single link.

Z. A. Azim, X. Fong, and K. Roy are with the School of Electrical and Computer Engineering, Purdue University, West Lafayette, IN 47907-1285 USA (email: zazim@purdue.edu).

T. Ostler, and R. Chantrell are with the Department of Physics, University of York, York YO10 5DD, UK

## II. LASER INDUCED MAGNETIZATION REVERSAL

It has been recently demonstrated, both theoretically and experimentally, that ultrafast heating using a femtosecond laser pulse can lead to deterministic magnetization reversal in Rare Earth (RE)-Transition Metal (TM) ferrimagnetic materials [4], [5]. The rapid transfer of thermal energy from the laser to the spin system leads to magnetization reversal within a few picoseconds and this mechanism is independent of the polarization of light for single-shot measurements [4]. The material system under consideration in this work is $Gd_{25}Fe_{65.4}Co_{9.6}$ in which the two sub-lattices, FeCo and Gd, are anti-ferromagnetically coupled in the ground state and point in anti-parallel direction (Fig. 1). When an ultrashort laser pulse excites GdFeCo, the electronic temperature increases sharply and creates a thermal bath for spins with temperature well above the Curie point. This causes a rapid energy transfer into the spin system which leads to the demagnetization of the sub-lattices but at vastly different rates due to their differing magnetic moments. Analysis of the spinwave spectra for GdFeCo shows that there are two branches in the dispersion [6]. The two branches can be thought of as one having ferromagnetic-like behavior and the other antiferromagnetic-like; however, as was shown in Ref. [6], these excitations are mixed. The switching was subsequently shown to occur when there is sufficient energy to excite spinwaves in both of these bands and can give rise to a transient ferromagnetic-like state [5]. After the initial heating, the electronic temperature quickly starts to come down towards equilibrium and as a result the Fe-Fe correlation starts to grow. As the Fe percentage is considerably higher than Gd percentage in the system, the growth in the Fe-Fe correlation means that there is a growing exchange on the Gd sub-lattice [5]. Consequently, the Gd sub-lattice reverses in the exchange field of the Fe, as visualized in Fig. 1. The non-equivalence of the sublattices in GdFeCo and the different demagnetization times of the magnetic species are essential for the process described here [4].

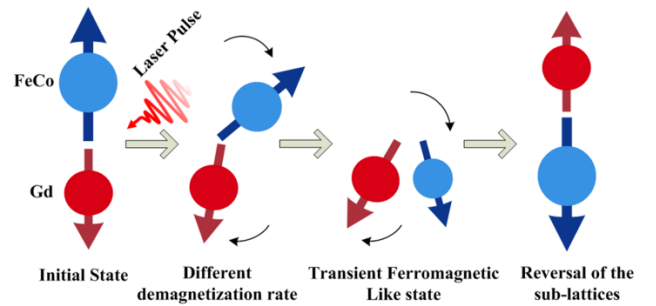

Fig.1. Switching mechanism for laser induced magnetization reversal.

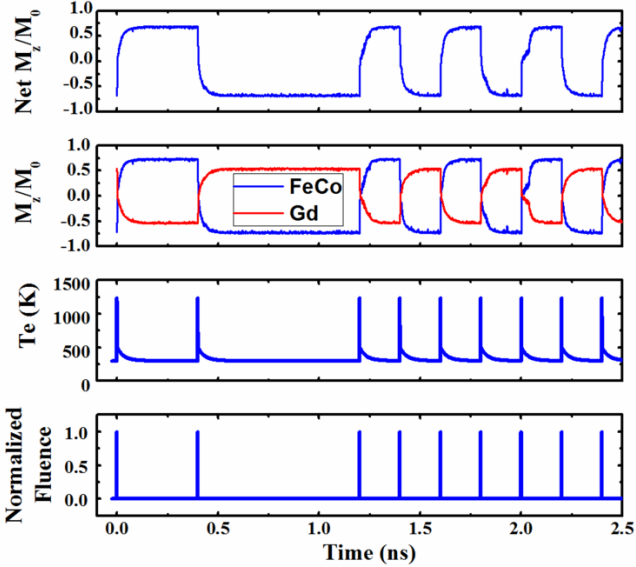

Fig.2. Response of $Gd_{25}Fe_{65.4}Co_{9.6}$ layer to a random sequence of ultrashort laser pulses.

In the present work, to model the ultrafast magnetization reversal process, we utilize the atomistic spin dynamics model [4]. This model treats the FeCo and Gd magnetic moments as localized to atomic sites. The exchange constants were determined through fitting to experimental observations [7]. The dynamic response of the system is simulated using Langevin Dynamics, essentially the Landau-Lifshitz-Gilbert equation augmented by a random field term to represent thermal effects [7]. After the action of a laser pulse; electronic temperature variation is calculated using the two-temperature model [8] and the spin system is coupled with the electronic system [4]-[6]. The electronic thermal bath is also coupled with the phonons and the thermal energy is removed by the phonons on the picosecond time-scale as the electronic system equilibrates with the surrounding external bath at room temperature (full details in Refs. [4] and [7]).

## III. Circuit Analysis and Discussion

In this work, we have analyzed a 9nm x 9nm x 3nm $Gd_{25}Fe_{65.4}Co_{9.6}$ layer, where the Gd percentage is chosen such that for the chosen fluence, the magnetization state will switch each time if it is excited by a laser pulse. Experimentally this threshold fluence is ~2.3 mJ/cm² [6]. The response of the chosen layer to a random sequence of laser pulses is shown in Fig. 2. The sub-lattices switch their magnetization after every 50fs pulse with energy density of 0.57 GJ/m³; hence the net magnetization switches also. The MTJ assumed in our design consists of a GdFeCo free layer and TbFeCo pinned layer with MgO as a tunneling barrier [9]. This structure (shown in Fig. 3) provides sharp magnetization reversal with a TMR of 64%. We perform NEGF based transport simulation [10] of the MTJ to investigate the effects of using this MTJ with CMOS circuits [11] and the overall simulation framework [12] is shown in Fig. 4. In our design, the MTJ is switched with the laser alone without the need for an external magnetic field. The optical modulator is designed such that, at each rising edge of the clock pulse, a 50fs pulse of linearly polarized laser is transmitted to excite the GdFeCo free layer if the data input is high. The laser fluence is kept above the switching value so that every time it strikes the free layer, the RE and TM sub-lattices switch their magnetization state and as a result the net magnetization switches as well. The state of the MTJ may be sensed using the voltage divider shown in Fig. 5. Full voltage swing is recovered using the inverters following the MTJ output voltage. In the receiver, flip-flops are used to check if the MTJ state has changed between two consecutive clock pulses. A change in the MTJ state results in a high signal on the output of the last flip-flop for that clock period. The overall circuit diagram is shown in Fig. 5. A simplified schematic of the laser focusing scheme using optical fiber link for this off-chip interconnect architecture is shown in Fig. 6. To verify the circuit functionality, we have simulated the effect of optically transmitting a single bit data stream. The results shown in Fig. 7 demonstrate that the receiver successfully recovers the transmitted signal.

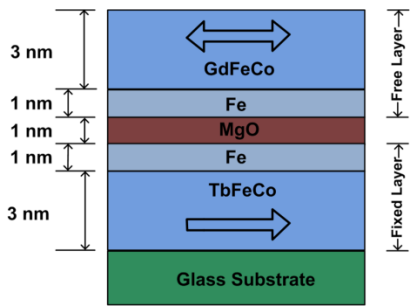

Fig.3. MTJ structure used in the design which has a TMR of 64%.

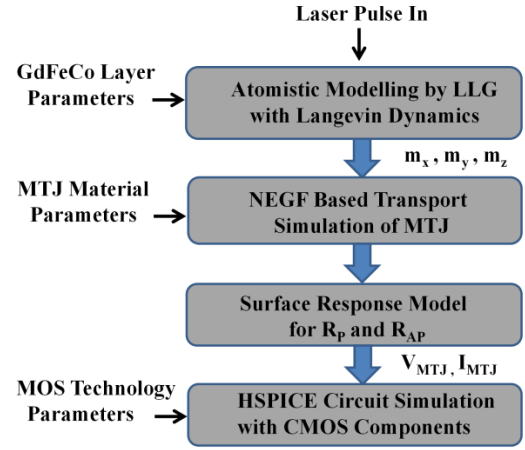

Fig.4. Device-Circuit Simulation framework used in this work.

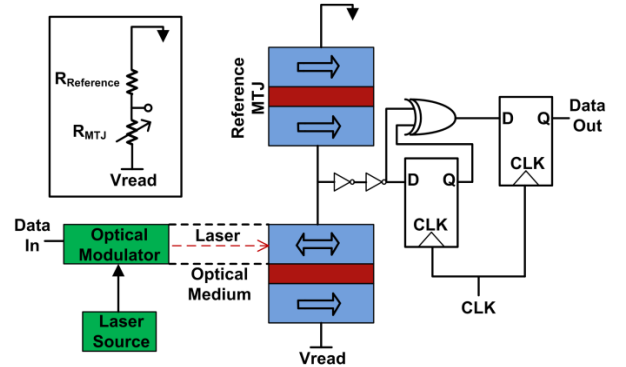

Fig.5. Detection circuit for optical interconnect using laser induced magnetization reversal. The MTJs act as a voltage divider shown inset.

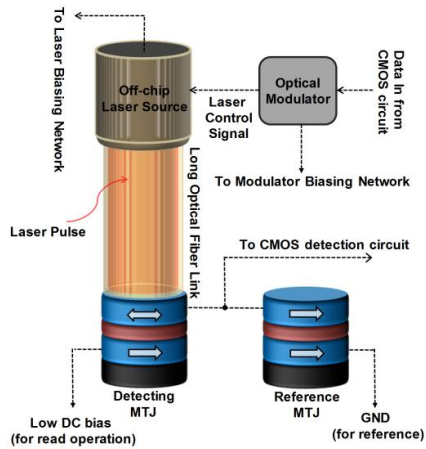

Fig.6. Schematic view of focusing of the laser beam on the detecting MTJ.

A significant portion of the energy dissipation in optical interconnects occurs in the photodetectors and subsequent amplifiers at the receiving side. In our proposed design, optical-to-electrical conversion is achieved by directly switching the MTJ using the laser alone. As a result there is no need to use photodetectors and amplifiers at the receiver. This offers tremendous improvement in performance and energy consumption as generating sufficient numbers of carriers in the photodetectors is very challenging. Total power dissipation in an optical interconnect scheme occurs as a summation in the optical modulator, optical channel, and the detector, where the consumption in the channel is negligible [2]. The energy consumption calculated for our scheme is 0.6 pJ/bit which includes the dissipation in both the standard modulator [2] and the CMOS detection circuitry. To make the computations numerically feasible, the size of the analyzed MTJ is taken to be relatively smaller than the actual size for physical implementation. The size of the MTJ can be increased to facilitate the focusing of the laser on nano-scale and there have been significant advancements in this field [13]. Increasing the MTJ size will have a negligible effect on the power consumption as the writing power is essentially coming from the laser and the reading power is mainly consumed in the CMOS components. So, increasing the size of the MTJ, even only the free layer of the MTJ to make the focusing of the laser feasible will have a relatively insensitive effect on the overall power consumption.

## IV. CONCLUSION

For a single link, our proposed interconnect can transmit data at 5 Gbits/sec while consuming 0.6 pJ/bit which follows the trend of the ITRS roadmap [14]. For the target speed, there is almost 40% improvement from the current technology value of ~1 pJ/bit and there have been continued efforts to achieve sub pJ/bit [14], [15]. Although, primarily designed to be used off-chip, this design has the potential to be extended for on-chip optical interconnects as the receiver design is made considerably simpler.

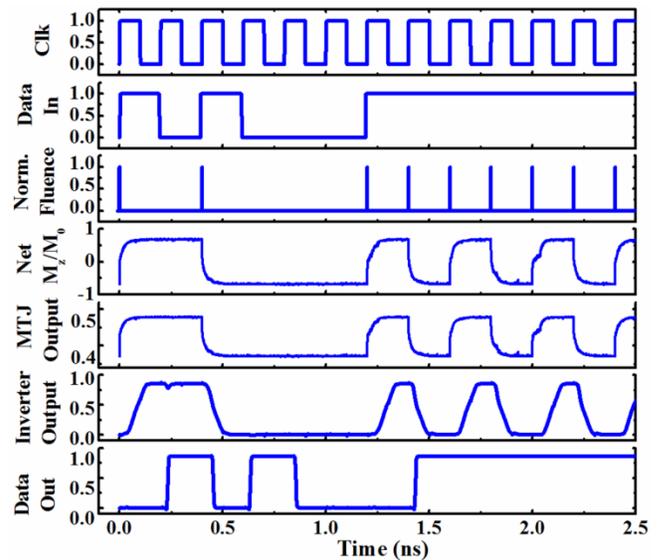

Fig.7. Operation of the detection circuit for a random input sequence.